\documentstyle[12pt]{article}

\author{Yu.~M.~Zinoviev
       \thanks{E-mail address: ZINOVIEV@MX.IHEP.SU} \\
        {\it Institute for High Energy Physics} \\
        {\it Protvino, Moscow Region, 142284, Russia}}
\title{On Massive High Spin Particles in (A)dS}
\date{}

\begin{document}

\maketitle

\begin{abstract}
  In this Letter we consider the problem of partial masslessness and
unitarity in (A)dS using gauge invariant description of massive high
spin particles. We show that for $S=2$ and $S=3$ cases such formalism
allows one correctly reproduce all known results. Then we construct a
gauge invariant formulation for massive particles of arbitrary
integer spin $s$ in arbitrary space-time dimension $d$. For $d=4$ our
results confirm the conjecture made recently by Deser and Waldron.
\end{abstract}

\newpage

\section*{Introduction}

Massive high spin particles in (A)dS reveal a number of interesting
and peculiar features, such as unitary forbidden mass range
and appearance of gauge invariance in partially massless theories \cite{DN83,DN84,Hig87,Ben95,DW01,DW01a,DW01b,DW01c}. In this Letter we
consider these properties using gauge invariant description of
massive high spin particles in constant curvature space-time
constructed in the same way as in the flat space case
\cite{Zin83,Zin94,KZ97}. Being unitary and gauge invariant from the
very beginning, such formalism appears to be very well suited for the
investigation of unitarity, gauge invariance and partial
masslessness. Moreover, such formalism could be useful for the
problem of van Dam-Veltman-Zakharov discontinuity
\cite{DDLS01,DLS01}.

We start with the $S=2$ and $S=3$ cases and show that all known
results concerning unitarity and partial masslessness can be
correctly reproduced and understood in our approach. Then we consider
massive particles of arbitrary integer spin $s$ in arbitrary
space-time dimension $d$. In particular, we calculate all critical
values for $m^2$ corresponding to partially massless theories. For
$d=4$ our results confirm the conjecture made recently by Deser and
Waldron \cite{DW01c}.

\section{Spin 2}

As a warming up exercise, let us start with the most simple and
rather well known case --- massive spin-2 particle. For the gauge
invariant description one needs three fields $(h_{\mu\nu}, A_\mu,
\varphi)$ where the first one is symmetric. We begin with the
Lagrangian:
\begin{eqnarray}
{\cal L}_0 &=& {\cal L}_0(h_{\mu\nu}) + {\cal L}_0(A_\mu) + {\cal
L}_0(\varphi) \\
{\cal L}_0(h_{\mu\nu}) &=& \frac{1}{2} D^\mu h^{\alpha\beta} D_\mu
h_{\alpha\beta} - (Dh)^\mu (Dh)_\mu - \frac{1}{2} D^\mu h D_\mu h -
(DDh) h     \nonumber \\
{\cal L}_0(A_\mu) &=& - \frac{1}{2} D^\mu A^\nu D_\mu A_\nu +
\frac{1}{2} (DA) (DA)  \qquad {\cal L}_0(\varphi) = \frac{1}{2} D^\mu
\varphi D_\mu \varphi  \nonumber
\end{eqnarray}
which is the sum of usual massless Lagrangians for these fields where
all derivatives are replaced by the covariant ones\footnote{Because
the covariant derivatives do not commute, there are slightly
different choices for their ordering in kinetic terms. As a result,
the structure of mass terms depends on the choice made.}. Here $(D
h)^\mu = D_\alpha h^{\alpha\mu}$, $(DDh) = D_\mu D_\nu h^{\mu\nu}$,
$h = g_{\mu\nu} h^{\mu\nu}$ and so on. Then we add all possible low
derivative terms:
\begin{eqnarray}
\Delta {\cal L} &=&   a_2 A^\mu (Dh)_\mu + b_2 h (DA) - a_1 \varphi
                       (DA)    \nonumber  \\
         & & {} + d_2 h^{\mu\nu} h_{\mu\nu} + e_2 h^2 + f_2 h
                \varphi - d_1 A_\mu{}^2 + d_0 \varphi^2 
\end{eqnarray}
and require that the whole Lagrangian be invariant under the
following gauge transformations:
\begin{eqnarray}
\delta h_{\mu\nu} &=& \frac{1}{2} (D_\mu \xi_\nu + D_\nu \xi_\mu) + 
\beta_2 g_{\mu\nu} \xi    \nonumber  \\
\delta A_\mu &=& D_\mu \xi + \alpha_1 \xi_\mu   \\
\delta \varphi &=&  \alpha_0 \xi     \nonumber
\end{eqnarray}
In this, we use the normalization
\begin{equation}
R_{\mu\nu,\alpha\beta} = - \Omega (g_{\mu\alpha} g_{\nu\beta} -
g_{\mu\beta} g_{\nu\alpha}), \qquad \Omega =
\frac{2\Lambda}{(d-1)(d-2)}
\end{equation}
The requirement of gauge invariance allows one to express all the
parameters in the Lagrangian as well as $\beta$
\begin{eqnarray*}
a_2 &=& - 2 \alpha_1 \qquad b_2 = - 2 \alpha_1 \qquad a_1 = -
\alpha_0 \qquad \beta_2 = \alpha_1 \\
d_2 &=& - \alpha_1{}^2 - \Omega \qquad e_2 = \alpha_1{}^2 -
\frac{\Omega}{2} \qquad f_2 = - \alpha_1 \alpha_0 \\
2d_1 &=& 6 \alpha_1{}^2 - \alpha_0{}^2 - 3\Omega \qquad d_0 = 2
\alpha_1{}^2
\end{eqnarray*}
and gives the relation on $\alpha$'s
\begin{equation}
\alpha_0{}^2 = 6 (\alpha_1{}^2 - \Omega)
\end{equation}
Here and further on, we will use the convention that $m$ is the
particle mass in the flat space limit $(\Lambda=0)$. In this case it
means
\begin{equation}
\alpha_1{}^2 = \frac{m^2}{2}
\end{equation}
and consequently
\begin{equation}
\alpha_0{}^2 = 3 (m^2 - 2\Omega)
\end{equation}
From the last relation it is evident that consistent description
exists in the unitary allowed region $(m^2 \ge 2\Omega)$ only. It is
easy to check that if one changes the sign of the scalar field
$\varphi$ kinetic term, then the gauge invariant formulation will be
possible for $m^2 < 2\Omega$, but unitarity will be lost the scalar
field being the ghost one.

For the critical value $(m^2 = 2\Omega)$ scalar field completely
decouples, while the rest fields $(h_{\mu\nu},A_\mu)$ describe
unitary partially massless theory with the helicities $(\pm 2,\pm
1)$. The Lagrangian in this case has a form:
\begin{eqnarray}
{\cal L} &=& {\cal L}_0(h_{\mu\nu}) + {\cal L}_0(A_\mu) - 2 M A^\mu
(Dh)_\mu - 2 M h (DA) \nonumber \\
 && {}  - 2 M^2 h^{\mu\nu}  h_{\mu\nu} + \frac{M^2}{2} h^2
 + \frac{3 M^2}{2} A_\mu{}^2
\end{eqnarray}
where $M^2 = \frac{\Lambda}{3}$ and it is invariant under the gauge
transformations:
\begin{eqnarray}
\delta h_{\mu\nu} &=& \frac{1}{2} (D_\mu \xi_\nu + D_\nu \xi_\mu) + M
g_{\mu\nu} \xi \nonumber \\
\delta A_\mu &=& D_\mu \xi + M \xi_\mu
\end{eqnarray}
Indeed we have $10+4=14$ independent components and $4+1=5$
gauge parameters leaving $14-2*5=4$ physical degrees of freedom.

As is well known, in flat space massive spin-2 particle in the
massless limit breaks into the massless particles with spins 2, 1 and
0. In AdS space the situation is different. If we consider $m
\rightarrow 0$ limit (which is consistent for $\Lambda < 0$ only) we
obtain massless spin-2 particle as well as massive spin-1 particle
described by the the fields ($A_\mu$, $\varphi$) with the Lagrangian:
\begin{equation}
{\cal L} = {\cal L}_0(A_\mu) + {\cal L}_0(\varphi) + M \varphi (DA) +
\frac{M^2}{4} A_\mu{}^2 \qquad M^2 = - 2 \Lambda
\end{equation}
which is invariant under gauge transformations:
\begin{equation}
\delta A_\mu = D_\mu \xi \qquad \delta \varphi = M \xi
\end{equation}

\section{Spin 3}

Let us turn to more interesting and instructive case of massive
spin-3 particles. For gauge invariant description of such particle
one needs four fields $\Phi_{\mu\nu\lambda}$, $h_{\mu\nu}$, $A_\mu$
and $\varphi$, the first two being symmetric ones. Again we start
with the sum of covariantized massless Lagrangians for all these
fields:
\begin{eqnarray}
{\cal L}_0 &=& {\cal L}_0(\Phi_{\mu\nu\lambda}) + {\cal
L}_0(h_{\mu\nu}) + {\cal L}_0(A_\mu) + {\cal L}_0(\varphi) \\
{\cal L}_0(\Phi_{\mu\nu\lambda}) &=& - \frac{1}{2} D^\mu
\Phi^{\nu\alpha\beta} D_\mu \Phi_{\nu\alpha\beta} + \frac{3}{2}
(D\Phi)^{\mu\nu} (D\Phi)_{\mu\nu} + \frac{3}{2} D^\mu
\widetilde{\Phi}^\nu D_\mu \widetilde{\Phi}_\nu \nonumber \\
 && {} + 3 (DD\Phi)^\mu \widetilde{\Phi}_\mu + \frac{3}{4} (D
 \widetilde{\Phi})  (D \widetilde{\Phi}) \nonumber
\end{eqnarray}
where Lagrangians for $h_{\mu\nu}$, $A_\mu$ and $\varphi$ are the
same as in the previous section, $\widetilde{\Phi}_\lambda =
g^{\mu\nu} \Phi_{\mu\nu\lambda}$, and add all possible low derivative
terms:
\begin{eqnarray}
\Delta {\cal L} &=& - a_3 h^{\mu\nu} (D \Phi)_{\mu\nu} - b_3
\widetilde{\Phi}^\mu (D h)_\mu - c_3 (D \widetilde{\Phi}) h
\nonumber  \\
 & & {} + a_2 A^\mu (D h)_\mu + b_2 h (D A) - a_1 \varphi (D A)
 \nonumber  \\
 & & {} - d_3 \Phi_{\mu\nu\lambda}{}^2 - e_3 \widetilde{\Phi}_\mu{}^2
 - f_3 \widetilde{\Phi}^\mu A_\mu + d_2 h_{\mu\nu}{}^2 \nonumber  \\
 & & {}  + e_2 h^2 + f_2 h \varphi - d_1 A_\mu{}^2  + d_0 \varphi^2
\end{eqnarray}
Then we require that the whole Lagrangian be invariant under the
following gauge transformations:
\begin{eqnarray}
\delta \Phi_{\mu\nu\lambda} &=& \frac{1}{3} (D_\mu \xi_{\nu\lambda}
+ D_\nu \xi_{\mu\lambda} + D_\lambda \xi_{\mu\nu}) + \beta_3
(g_{\mu\nu} \xi_\lambda + g_{\mu\lambda} \xi_\nu + g_{\nu\lambda}
\xi_\mu)       \nonumber  \\
\delta h_{\mu\nu} &=&  \frac{1}{2} (D_\mu \xi_\nu + D_\nu \xi_\mu) +
\beta_2 g_{\mu\nu} \xi + \alpha_2 \xi_{\mu\nu}      \nonumber  \\
\delta A_\mu &=&  D_\mu \xi + \alpha_1 \xi_\mu         \\
\delta \varphi &=& \alpha_0 \xi      \nonumber
\end{eqnarray}
where parameter $\xi_{\mu\nu}$ is symmetric and traceless. The
requirement of gauge invariance allows one to express all the
parameters in the Lagrangian as well as $\beta_{2,3}$
\begin{eqnarray*}
a_3 &=& - 3 \alpha_2 \qquad b_3 = - 6 \alpha_2 \qquad c_3 = -
\frac{3}{2} \alpha_2 \qquad \beta_3 = \frac{\alpha_2}{4} \\
a_2 &=& - 2 \alpha_1 \qquad b_2 = - 2 \alpha_1 \qquad a_1 = -
\alpha_0 \qquad \beta_2 = \alpha_1 \\
2d_3 &=& - 3 \alpha_2{}^2 + \Omega \qquad 2e_3 = 9 \alpha_2{}^2 - 18
\Omega \qquad f_3 = - 3 \alpha_2 \alpha_1 \\
2d_2 &=& - 2 \alpha_1{}^2 + \frac{15}{2} \alpha_2{}^2 - 2 \Omega
\qquad 2e_2 = 2 \alpha_1{}^2 - 3 \alpha_2{}^2 - \Omega \qquad f_2 = -
\alpha_1 \alpha_0 \\
2d_1 &=& - \alpha_0{}^2 + 6 \alpha_1{}^2 - 3 \Omega \qquad d_0 = 2
\alpha_1{}^2
\end{eqnarray*}
and gives two relations on $\alpha$'s
\begin{eqnarray*}
\alpha_1{}^2 &=& \frac{15}{4} \alpha_2{}^2 - 5 \Omega \\
\alpha_0{}^2 &=& - \frac{9}{2} \alpha_2{}^2 + 6 \alpha_1{}^2 - 6
\Omega
\end{eqnarray*}
Our convention that $m$ is the particle mass in the flat space gives
now $\alpha_2{}^2 = m^2/3$ and one finds:
\begin{eqnarray*}
\alpha_1{}^2 &=& \frac{5}{4} (m^2 - 4\Omega) \\
\alpha_0{}^2 &=& 6 (m^2 - 6 \Omega)
\end{eqnarray*}
One can see that now we have two critical values for the cosmological
constant. At the first one ($m^2 = 6 \Omega$) scalar field $\varphi$
completely decouples, while the rest fields describe unitary
partially massless theory with helicities $\pm 3,\pm 2,\pm 1$.
Indeed, we have total $20 + 10 + 4 = 34$ independent field components
and $9 + 4 + 1 = 14$ gauge parameters which gives $34 - 2*14 = 6$
physical degrees of freedom. In general for $m^2 < 6 \Omega$ the
theory becomes inconsistent, but at the second critical value $m^2 =
4 \Omega$ one obtains two completely decoupled systems. The one with
fields $\Phi_{\mu\nu\lambda}$, $h_{\mu\nu}$ describes unitary
partially massless theory with helicities $\pm 3, \pm 2$. The
Lagrangian for these fields has the form:
\begin{eqnarray}
{\cal L} &=& {\cal L}_0(\Phi_{\mu\nu\lambda}) + {\cal
L}_0(h_{\mu\nu}) + 3 M h^{\mu\nu} (D\Phi)_{\mu\nu} + 6 M
\widetilde{\Phi}^\mu (Dh)_\mu + \frac{3M}{2} (D\widetilde{\Phi}) h
\nonumber \\
 && {} + \frac{9M^2}{8} \Phi_{\mu\nu\lambda}{}^2 + \frac{9M^2}{4}
 \widetilde{\Phi}_\mu{}^2 + 3 M^2 h_{\mu\nu}{}^2 - \frac{15 M^2}{8}
 h^2
\end{eqnarray}
and corresponding gauge transformations are:
\begin{eqnarray}
\delta \Phi_{\mu\nu\lambda} &=& \frac{1}{3} (D_\mu \xi_{\nu\lambda}
+ D_\nu \xi_{\mu\lambda} + D_\lambda \xi_{\mu\nu}) + \frac{M}{4}
(g_{\mu\nu} \xi_\lambda + g_{\mu\lambda} \xi_\nu + g_{\nu\lambda}
\xi_\mu)       \nonumber  \\
\delta h_{\mu\nu} &=&  \frac{1}{2} (D_\mu \xi_\nu + D_\nu \xi_\mu)
 + M \xi_{\mu\nu}
\end{eqnarray}
Thus one has $20 + 10 = 30$ independent components, $9 + 4 = 13$
gauge parameters and $30 - 2*13 = 4$ physical degrees of freedom.

If we consider massless $m \rightarrow 0$ limit in AdS space $\Lambda
< 0$ we obtain massless spin-3 particle as well as the system of
($h_{\mu\nu}$, $A_\mu$, $\varphi$) fields giving gauge invariant
description of massive spin-2 particle exactly as in the previous
section.

\section{Arbitrary integer spin}

Now we are ready to consider general case of massive particle with
arbitrary integer spin $s$ in arbitrary space-time dimension $d$. In
order to avoid complex expressions with indices, we will use
condensed notations \cite{KZ97}. Let us denote $\Phi^s$ completely
symmetric tensor of rank $s$, which is double traceless
$\widetilde{\widetilde{\Phi}} = 0$, where $\widetilde{\Phi} =
Tr(\Phi^s)$, $\widetilde{\widetilde{\Phi}} = Tr(Tr(\Phi^s))$ and so
on, $Tr$ is a contraction of two indices by metric tensor. For the
description of massless particles with spin $s$ we will use the
Lagrangian \cite{Fro78}:
\begin{eqnarray*}
  {\cal L}_o(\Phi^s) &\!=\!& (-1)^s \biggl\{ \frac{1}{2}
  (\partial_\mu \Phi^s)  (\partial_\mu \Phi^s) - \frac{s}{2}
  (\partial \cdot \Phi^s)   (\partial \cdot \Phi^s) -
  \frac{s(s-1)}{4} (\partial_\mu  \widetilde{\Phi}^s)
  (\partial_\mu \widetilde{\Phi}^s)  \\
  && {} - \frac{s(s-1)}{2} (\partial \cdot \partial \cdot \Phi^s)
  \widetilde{\Phi}^s - \frac{1}{8} s(s-1)(s-2) (\partial \cdot
  \widetilde{\Phi}^s) (\partial \cdot \widetilde{\Phi}^s) \biggr\}
\end{eqnarray*}
where $(\partial \cdot \Phi^s) \stackrel{def}{=} \partial_\mu
\Phi^{\mu\mu_2\ldots\mu_s}$, $(\partial \cdot \partial \cdot \Phi^s)
\stackrel{def}{=} \partial_\mu \partial_\nu 
\Phi^{\mu\nu\mu_3\ldots\mu_s}$, and point denotes contraction of all
indices between tensor objects, for example, $\Phi^s \cdot \Phi^s
\stackrel{def}{=} \Phi_{\mu_1\ldots\mu_s} \Phi^{\mu_1\ldots\mu_s}$.
In flat space this Lagrangian is invariant under the following gauge
transformations:
\begin{equation}
\delta \Phi^s = \frac{1}{s} \cdot \left\{ \partial \xi^{s-1}
\right\}_{s.}
\end{equation}
where $\{\ldots\}_{s.}$ means symmetrization over all indices
(without normalization) and $\xi^{s-1}$ is symmetric traceless
tensor of rank $s-1$.

To obtain a description for massive particle we start with the sum of
massless Lagrangians for fields $(\Phi^s, \Phi^{s-1},...,\Phi^1,
\Phi^0)$:
\begin{equation}
  {\cal L}_0 = \sum_{k=0}^s {\cal L}_0(\Phi^k)
\end{equation}
where all derivatives now are covariant and add the most general low
derivative terms:
\begin{eqnarray}
\Delta {\cal L} &=& \sum_k (-1)^k \left[ a_k \Phi^{k-1} (D \cdot
\Phi^k) + b_k \widetilde{\Phi}^k (D \Phi^{k-1}) + c_k (D \cdot
\widetilde{\Phi}^k) \widetilde{\Phi}^{k-1} \right. \nonumber \\
 && {} \qquad \qquad \left. + d_k (\Phi^k)^2 + e_k
 (\widetilde{\Phi}^k)^2 +  f_k  \widetilde{\Phi}^k \Phi^{k-2} \right]
\end{eqnarray}
Then we require that the whole Lagrangian be invariant under the
following gauge transformations:
\begin{equation}
\delta \Phi^k = \frac{1}{k} \left\{ \partial \xi^{k-1} \right\}_{s.}
+ \alpha_k \xi^k + \beta_k \left\{ g^2 \xi^{k-2} \right\}_{s.}
\end{equation}
where the first term is absent at $k=0$ and the third one at $k<2$,
while $g^2$ is a metric tensor. This allows one to express all the
parameters in the Lagrangian as well as $\beta$'s:
\begin{eqnarray*}
a_k &=& - k \alpha_{k-1} \qquad b_k = - k(k-1) \alpha_{k-1} \qquad
c_k = - \frac{k(k-1)(k-2)}{4} \alpha_{k-1}  \\
2d_k &=& \frac{2(k+1)(2k+d-3)}{(2k+d-4)} \alpha_k{}^2 - k
\alpha_{k-1}{}^2 + \Omega [(k-1)(k-4) + (k-2)(d-1)], \quad k \ge 1 \\
d_0 &=& \frac{d}{d-2} \alpha_1{}^2 \qquad \beta_k = \frac{2
\alpha_{k-1}}{(k-1)(2k+d-6)}  \\
2e_k &=& - \frac{k(k^2-1)(2k+d)}{4(2k+d-4)} \alpha_k{}^2 +
\frac{k^2(k-1)}{2} \alpha_{k-1}{}^2 - \Omega \frac{k(k-1)}{2}[k(k-3)
+ (k-1)(d-1)] \\
2f_k &=& - k(k-1) \alpha_{k-1} \alpha_{k-2}
\end{eqnarray*}
and gives recurrent relation on $\alpha$'s:
\begin{equation}
(k-1) \alpha_{k-2}{}^2 = - \frac{(k+1)(2k+d-2)}{2k+d-4)} \alpha_k{}^2
+ \frac{2k(2k+d-5)}{(2k+d-6)} \alpha_{k-1}{}^2 - 2\Omega (2k+d-5),
\qquad k \ge 2
\end{equation}
In order to solve this relation we use our usual convention on mass
normalization which in this case gives $\alpha_{s-1}{}^2 = m^2/s$. We
get:
\begin{equation}
\alpha_k{}^2 = \frac{(s-k)(s+k+d-3)}{(k+1)(2k+d-2)} [m^2 - \Omega
(s-k-1)(s+k+d-4)], \qquad 0 \le k \le s-2
\end{equation}
In some sense the last formula is the main result of our work. It
gives all the critical values for the cosmological constant at which
partial masslessness appears\footnote{After this paper was finished and sent to electronic archive, we became aware about the work of A.~Higuchi \cite{Hig87a}, where these critical values were obtained by use of group theory methods.}. Indeed, it's easy to see that then,
say, $\alpha_k = 0$ the whole system decompose onto two subsystems,
one of them (with the fields $\Phi^s,\Phi^{s-1},...,\Phi^{k+1}$)
describing unitary partially massless theory. For $d=4$ our results
coincide with the conjecture made recently by Deser and Waldron
\cite{DW01c}.

\vspace{0.3in}
{
\centerline{\large \bf Acknowledgments}
}
\vspace{0.2in}

I am grateful to S.~M.~Klishevich for collaboration on earlier stages
of this work.

\end{document}